\pdfoutput=1
\documentclass[10pt,twocolumn,letterpaper]{article}
\usepackage[letterpaper,margin=0.75in,columnsep=0.25in]{geometry}
\usepackage{graphicx}
\usepackage{booktabs}
\usepackage{balance}
\usepackage{amsmath,amssymb,bm}
\usepackage{cite}
\usepackage[font=small,labelfont=bf,labelsep=period]{caption}

\usepackage{placeins}
\usepackage{titlesec}
\titleformat{\section}{\normalfont\normalsize\bfseries}{\thesection.}{0.5em}{}
\titlespacing*{\section}{0pt}{1.8ex plus 0.3ex minus 0.2ex}{0.8ex plus 0.2ex}
\titleformat{\subsection}{\normalfont\normalsize\itshape}{\thesubsection.}{0.5em}{}
\titlespacing*{\subsection}{0pt}{1.4ex plus 0.2ex minus 0.2ex}{0.5ex plus 0.1ex}
\usepackage[colorlinks=true,allcolors=blue]{hyperref}
\hypersetup{pdftitle={Cross-environment diagnostics for new physics in proton-proton and heavy-ion collisions},pdfauthor={Yi Yang},pdfsubject={Unresolved contributions to resonance signals across collision systems},pdfkeywords={new physics,quarkonium,nuclear modification,heavy-ion collisions}}
\newcommand{\Q}{\mathcal Q}
\newcommand{\papertitle}{Cross-environment diagnostics for new physics in proton-proton and heavy-ion collisions}
\newcommand{\paperabstract}{Particle-physics searches and heavy-ion measurements probe many of the same final states, but their results are often interpreted separately. We propose testing new contributions hidden beneath established resonances by requiring a common candidate hypothesis across collision systems. Different nuclear responses change the relative populations of unresolved components and link proton-proton production and mass distributions to nuclear modification and azimuthal anisotropy. We derive these relations and, for component mass distributions of the same shape within each system, a relation between total nuclear modification and the calibrated shift in the mean invariant mass in which quarkonium yield suppression cancels. The candidate component's nuclear modification remains an input to this test. An example near the $\Upsilon(1S)$ follows a component absorbed into proton-proton fits into nuclear observables; for the specified responses, its fraction rises from five percent in proton-proton collisions to thirteen percent in nucleus--nucleus collisions. These results provide conditional consistency tests linking particle and nuclear measurements. Their use requires constrained candidate responses and controlled quarkonium and detector uncertainties; the statistical gain from adding nuclear data must be evaluated with the actual event samples.}
\begin{document}
\twocolumn[{
\centering
{\Large\bfseries\papertitle\par}
\vspace{0.7em}
{\normalsize Yi Yang$^{1,2}$\par}
\vspace{0.4em}
{\small $^1$Institute of Physics, Academia Sinica, Taipei 11529, Taiwan\par
$^2$Department of Physics, National Cheng Kung University, Tainan 70101, Taiwan\par
Email: yiyang429@as.edu.tw\par}
\vspace{0.8em}
\begin{minipage}{0.96\textwidth}\small
\textbf{Abstract}\par\smallskip
\paperabstract
\end{minipage}
\vspace{1.1em}
}]

\section{Introduction}
Particle-physics searches and heavy-ion studies often address different questions, even when they measure the same final state. Yet their interpretations are closely linked: related production processes contribute in each system, and proton-proton ($pp$) measurements provide the reference for proton--nucleus ($pA$) and nucleus--nucleus ($AA$) observables \cite{Andronic2016}. Bringing these measurements together can provide clues to new physics that separate interpretations may miss.

An additional contribution beneath a known resonance can shift its fitted yield, momentum spectrum, or line shape without producing a separate peak. If it is absorbed into a $pp$ reference, it also enters the inferred nuclear response. Related effects arise when new physics contaminates a control region or a precision-measurement sample \cite{Feigl2012,Agashe2024}. Dimuon searches in the bottomonium region already confront the difficulty of overlapping resonances \cite{HaischKamenik2016,LHCbDimuon2018}. The LHCb analysis omits mass hypotheses whose fitted candidate yield is strongly correlated with an $\Upsilon$ yield \cite{LHCbDimuon2018}.

The suppression of quarkonium (heavy-quark--antiquark bound states) is an established probe of deconfined QCD matter \cite{MatsuiSatz1986}. An additional source of the same dilepton final state need not have the same nuclear response. Their relative populations can therefore change even when the peaks remain unresolved. We propose testing the same candidate contribution jointly in $pp$, $pA$, and $AA$ collisions. A common candidate hypothesis links its production, decay, and nuclear interactions to nuclear yields, mass moments, and anisotropy.

Previous heavy-ion proposals for new-physics searches exploit enhanced photon fluxes, low pileup, special triggers, or ion-specific production mechanisms \cite{KnapenALP,BruceHIBSM,DrewesHI,YangLinALP}. Quarkonium supplies measured production, polarization, nuclear modification, and azimuthal anisotropy across collision systems \cite{CMSUpsilonPPPol,CMSUpsilonRpPb,CMSUpsilonRAA,CMSUpsilonV2}. Gambhir et al.\ used anomaly detection and likelihood-ratio event weights to identify anti-isolated $\Upsilon$ decays within CMS $pp$ data \cite{GambhirUpsilon}. Here nuclear response changes the composition of an existing signal, linking particle and nuclear measurements.

\section{Signal composition}
Consider incoherent contributions from a known quarkonium state $\Q$ and a narrow nearby component $X$. Throughout, $s$ labels the collision system ($pp$, $pA$, or $AA$), and $j$ labels $\Q$, $X$, or their total signal, denoted by $\mathrm{tot}$. At collision energy per nucleon pair $\sqrt{s_{NN}}$, $N_{j,s}$ is the yield of $j$ in a specified $(p_T,y)$ bin, where $y$ is measured in the nucleon--nucleon centre-of-mass frame. The yields use the same fiducial definition, with acceptance and efficiency corrections applied to each component. Nuclear observables refer to a specified event class, for example an $AA$ centrality interval characterizing the nuclear overlap.

The total signal excludes the background and is
$N_{\mathrm{tot},s}=N_{\Q,s}+N_{X,s}$.
Define the $pp$ yield ratio by
\[
r_{pp}(p_T,y;\sqrt{s_{NN}})
=\frac{N_{X,pp}}{N_{\Q,pp}},
\]
and the fraction of $X$ by $f_s=N_{X,s}/N_{\mathrm{tot},s}$,
so that $f_{pp}=r_{pp}/(1+r_{pp})$.
The mass splitting is $\Delta m=m_{\Q}-m_X$.
Here closure means consistency of the measurements across systems under the same candidate hypothesis.

Each nuclear sample and its $pp$ reference use the same collision energy and kinematic interval; we suppress these arguments below. For integrated observables, all yields and cross sections use the same region, and $r_{pp}$ is the ratio of integrated component cross sections. A ratio at one momentum does not define this integrated ratio. Different $p_T$ or rapidity selections can change the composition even for equal nuclear responses. Data at different energies require corresponding $pp$ references and a consistent energy dependence of both component spectra.

For $s=pA$ or $AA$, the nuclear modification factor is
\begin{equation}
R_s^j=\frac{1}{\langle T_s\rangle}
\frac{d^2N_{j,s}/(dp_T\,dy)}{d^2\sigma_{j,pp}/(dp_T\,dy)},
\label{eq:r_def}
\end{equation}
where $d^2N_{j,s}/(dp_T\,dy)$ is the nuclear yield per event. The average overlap function $\langle T_s\rangle$, with units of inverse cross section, converts the $pp$ cross section to the expected per-event yield from independent nucleon--nucleon collisions. Thus $R_s^j=1$ denotes binary-collision scaling, $R_s^j<1$ suppression, and $R_s^j>1$ enhancement. Both initial-state nuclear effects and final-state medium interactions can modify $R_s^j$ \cite{Andronic2016}.

Both components enter the nuclear yield and the $pp$ reference. The total signal therefore has
\begin{equation}
R_s^{\mathrm{tot}}=
\frac{R_s^{\Q}+r_{pp}R_s^X}
{1+r_{pp}}.
\label{eq:r_mix}
\end{equation}
This averages the two nuclear responses with their $pp$ fractions. The same $r_{pp}$ applies to both nuclear systems at fixed energy and kinematics.

Elliptic flow is quantified by the second Fourier coefficient of the signal's azimuthal distribution. Direction-dependent interactions in the anisotropic overlap region of noncentral $AA$ collisions can generate such an anisotropy \cite{CMSUpsilonV2}. Define \mbox{$v_2^j=\langle\cos2(\phi-\Psi_2)\rangle_j$}, where $\phi$ is the azimuth of the dilepton momentum about the beam and $\Psi_2$ specifies the second-harmonic symmetry plane. Positive $v_2^j$ favors emission along this plane over the perpendicular direction. The component coefficients use the same event class, reference plane, and correction procedure. Adding their azimuthal distributions gives the total coefficient in $AA$,
\begin{equation}
v_2^{\mathrm{tot}}=
\frac{R_{AA}^{\Q}v_2^{\Q}+r_{pp}R_{AA}^{X}v_2^X}
{R_{AA}^{\Q}+r_{pp}R_{AA}^{X}}.
\label{eq:v_mix}
\end{equation}
For $v_2^X=0$, the added component dilutes a nonzero $v_2^{\Q}$ because it contributes to the total yield but not to its second harmonic.

The nuclear fraction of $X$ is then
\begin{equation}
f_s=\frac{r_{pp}R_s^X}{R_s^{\Q}+r_{pp}R_s^X},
\qquad
\frac{f_s}{1-f_s}
=\frac{f_{pp}}{1-f_{pp}}\frac{R_s^X}{R_s^{\Q}}.
\label{eq:fraction}
\end{equation}
A component less suppressed than quarkonium is enriched in the surviving signal. This changes both the nuclear modification factor and, if $v_2^X$ differs from $v_2^{\Q}$, the observed anisotropy.

If all component nuclear responses and anisotropies are free, yields and flow do not determine $r_{pp}$. Testing a common candidate therefore requires predictions or independent constraints on its response and control of quarkonium and detector effects.

\section{Mass moments and nuclear modification}
The calibrated mean invariant mass supplies a test independent of quarkonium yield suppression. Within each system, assume that the normalized component mass profiles, including detector smearing, have the same shape and differ only by a translation,
\begin{equation}
S_{j,s}(m)=g_s(m-m_j),\qquad j\in\{\Q,X\},
\label{eq:response}
\end{equation}
with finite first moment and masses fixed across systems. The shape $g_s$, including its width, tails, and calibration offset, may differ between collision systems.

Define $\bar m_{\mathrm{tot},s}$ as the mean of the corrected total signal and $\bar m_{\Q,s}$ as the mean of an independently calibrated pure-$\Q$ response. The calibration must not use the possibly mixed sample's single-component fit, which can absorb the displacement. The translated profiles give
\begin{equation}
\delta_s\equiv\bar m_{\Q,s}-\bar m_{\mathrm{tot},s}=f_s\Delta m.
\label{eq:moment}
\end{equation}
For $X$ below quarkonium, $\delta_s>0$ describes a downward shift in the mean mass. This quantity is a signal first moment, rather than the location parameter of a single-peak fit.

Eqs.~\eqref{eq:r_mix} and \eqref{eq:fraction} give
\begin{equation}
R_s^{\mathrm{tot}}f_s
=\frac{R_s^{\Q}+r_{pp}R_s^X}{1+r_{pp}}
 \frac{r_{pp}R_s^X}{R_s^{\Q}+r_{pp}R_s^X}
=f_{pp}R_s^X.
\label{eq:fraction_identity}
\end{equation}
Multiplying by $\Delta m$ yields
\begin{equation}
R_s^{\mathrm{tot}}\,\delta_s=R_s^X\,\delta_{pp}.
\label{eq:moment_yield}
\end{equation}
The quarkonium yield suppression cancels, but $R_s^X$ remains. Equation~\eqref{eq:moment_yield} therefore tests a predicted or independently constrained candidate response; binary-collision scaling, $R_s^X=1$, predicts $R_s^{\mathrm{tot}}\delta_s=\delta_{pp}$. With unrestricted $R_s^X$ and $\delta_{pp}\ne0$, it instead infers $R_s^X=R_s^{\mathrm{tot}}\delta_s/\delta_{pp}$ within the mixture hypothesis; it does not establish $X$.

\begin{table*}[t]
\centering
\caption{Numerical inputs and spectrum-fit parameters for the quarkonium example. The spectrum normalization is in arbitrary units. Signal counts in the mass example refer to the full response functions; background counts refer to the fit window. The mass scan uses the same tail and background shapes as the fixed mass example.}
\label{tab:inputs}
\small
\setlength{\tabcolsep}{7pt}
\renewcommand{\arraystretch}{1.12}
\begin{tabular}{@{}p{0.18\textwidth}p{0.39\textwidth}p{0.36\textwidth}@{}}
\toprule
Calculation & Quantity & Value \\
\midrule
$pp$ spectrum & Quarkonium shape $(A_{\Q},p_{0,\Q},n_{\Q})$ & $(4.0\times10^6,\ 4.8~\mathrm{GeV},\ 5.65)$ \\
 & Additional shape $(p_{0,X},n_X)$ & $(6.0~\mathrm{GeV},\ 5.00)$ \\
 & Reference fraction & $f_{pp}=5\%$ at $p_T=20$~GeV \\
 & Momentum interval & $6\leq p_T\leq55$~GeV \\
 & Relative uncertainties & $2.8\%$ to $8.2\%$, linear in $p_T$ \\
 & Single-component fit $(A_{\rm fit},p_{0,\rm fit},n_{\rm fit})$ & $(4.14\times10^6,\ 4.70~\mathrm{GeV},\ 5.58)$ \\
\midrule
$pp$ mass fit & Quarkonium mass $m_{\Q}$ & $9.4603$~GeV \\
 & Mass splitting $\Delta m$; resolution $\sigma_m$ & $+25$~MeV; $70$~MeV \\
 & Crystal Ball tails $(\alpha,n_{\rm CB})$ & $(1.5,\ 5)$ \\
 & Signal counts $(N_{\Q,pp},N_{X,pp})$ & $(9500,\ 500)$ \\
 & Background count $N_{\rm bkg}$; slope $\beta$ & $6200$; $2.1$~GeV$^{-1}$ \\
 & Mass window; bin width & $9.18<m_{\mu\mu}<9.72$~GeV; $6$~MeV \\
\midrule
Mass scan & Signal count; background count; fraction & $N_{\Q,pp}=12000$; $N_{\rm bkg}=7000$; $f_{pp}=5\%$ \\
\midrule
Nuclear response & Quarkonium $(R_{pA}^{\Q},R_{AA}^{\Q})$ & $(0.90,\ 0.35)$ \\
 & Additional component $(R_{pA}^{X},R_{AA}^{X})$ & $(1,\ 1)$ \\
 & Elliptic flow $(v_2^{\Q},v_2^X)$ & $(0.030,\ 0)$ \\
\bottomrule
\end{tabular}
\end{table*}

\begin{figure}[t]
\centering
\includegraphics[width=0.98\columnwidth]{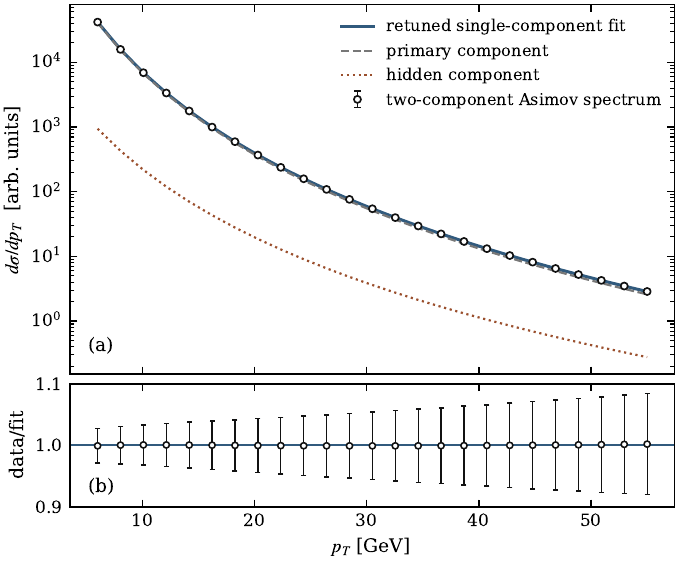}
\caption{Expected $pp$ spectrum containing quarkonium and a harder contribution with $f_{pp}=5\%$ at $p_T=20$~GeV. The sum is fitted by a single power law with free normalization, scale, and exponent. The lower panel shows the ratio to the fit. Points are expected values without statistical fluctuations.}
\label{fig:pp_spectrum}
\end{figure}

\begin{figure}[t]
\centering
\includegraphics[width=0.98\columnwidth]{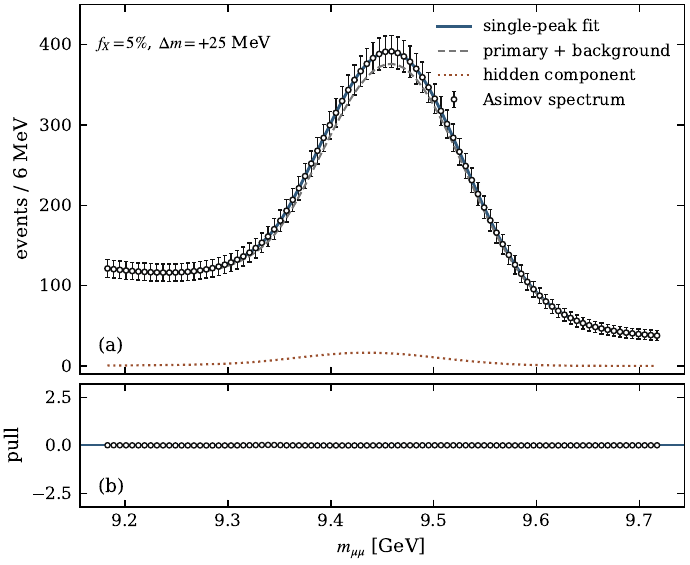}
\caption{Expected dimuon mass spectrum with a five-percent component $25$~MeV below the $\Upsilon(1S)$. The solid curve is a single-peak Crystal Ball fit with an exponential background. The dashed and dotted curves show quarkonium plus background and the additional contribution, respectively.}
\label{fig:mass_spectrum}
\end{figure}

\begin{figure}[t]
\centering
\includegraphics[width=0.98\columnwidth]{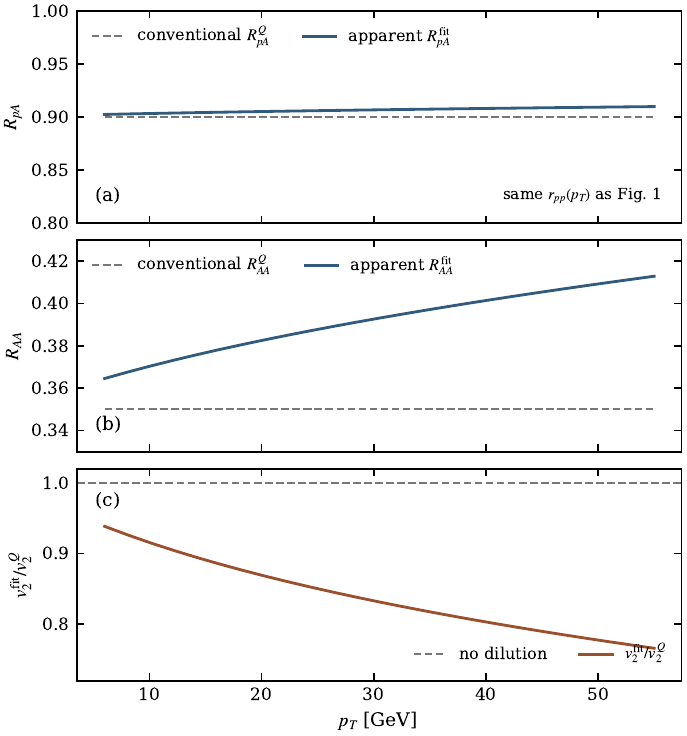}
\caption{Nuclear observables for the component in Fig.~\ref{fig:pp_spectrum}. Panels (a) and (b) show $R_{pA}^{\mathrm{tot}}$ and $R_{AA}^{\mathrm{tot}}$ for the quarkonium responses in Table~\ref{tab:inputs}. Panel (c) shows $v_2^{\mathrm{tot}}/v_2^{\Q}$. The additional component has $R_{pA}^{X}=R_{AA}^{X}=1$ and $v_2^X=0$.}
\label{fig:medium_response}
\end{figure}

\begin{figure}[t]
\centering
\includegraphics[width=0.98\columnwidth]{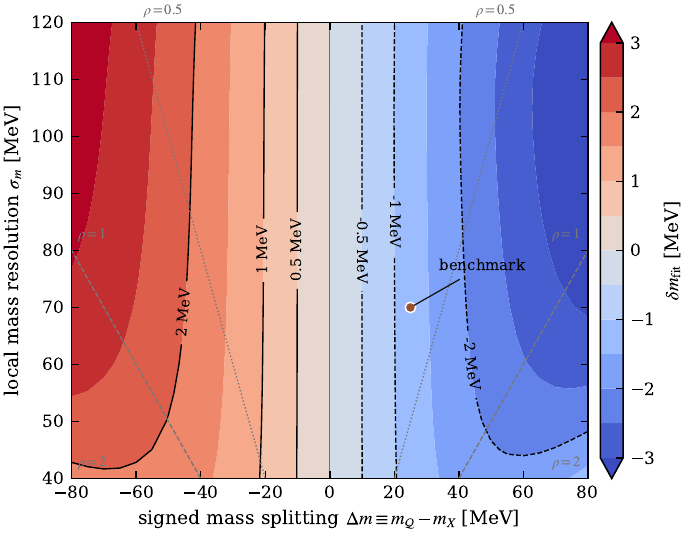}
\caption{Shift $\delta m_{\rm fit}$ of the single-peak fit location for a five-percent component, as a function of signed mass splitting and resolution. Gray lines mark selected values of $|\Delta m|/\sigma_m$. The marked point is the mass example in Table~\ref{tab:inputs}.}
\label{fig:mass_contours}
\end{figure}

Effects that change only the quarkonium yield, including energy loss or feed-down from heavier quarkonia, cancel from Eq.~\eqref{eq:moment_yield}. Quarkonium mass-profile changes, calibration offsets, background subtraction, and mass-dependent acceptance require separate control. Finite-window moments need window corrections. A fitted peak position requires a calibrated relation to the signal first moment.

\section{Quarkonium illustration}
We use expected distributions and fixed nuclear responses to follow one additional dimuon component from $pp$ fits to nuclear observables. It has a harder transverse-momentum spectrum than the $\Upsilon(1S)$, so its fraction grows with $p_T$. The numerical settings and spectrum-fit parameters are collected in Table~\ref{tab:inputs}.

\subsection{Absorption in proton-proton fits}
The individual $pp$ production spectra, for $j\in\{\Q,X\}$, are parametrized as
\begin{equation}
P_{j,pp}(p_T)=A_j\left(1+\frac{p_T}{p_{0,j}}\right)^{-n_j}.
\label{eq:pp_spectrum}
\end{equation}
The normalization of $X$ gives a five-percent fraction at the reference momentum. Its fraction rises from about two percent to ten percent over the momentum interval in Table~\ref{tab:inputs}.

A single power law fits the summed expected spectrum in Fig.~\ref{fig:pp_spectrum}. The fitted normalization, momentum scale, and exponent shift to accommodate the additional contribution. Thus the production fit alone can absorb the component into the inferred quarkonium spectrum.

At the reference momentum, the same component can also be hidden in the dimuon mass distribution,
\begin{equation}
\begin{aligned}
F_{pp}(m)={}&N_{\Q,pp}S_{\Q,pp}(m)\\
&+N_{X,pp}S_{X,pp}(m)+B(m).
\end{aligned}
\label{eq:mass_response}
\end{equation}
Both signals have Crystal Ball responses, with a Gaussian core joined to a low-mass power-law tail, while the background is exponential:
$B(m)\propto\exp[-\beta(m-m_{\min})]$,
where $m_{\min}$ is the lower edge of the fit window.

We fit the expected bin counts with one Crystal Ball signal and an exponential background using weighted least squares; each variance is the input expected count. The signal and background normalizations, signal location, width, $\alpha$, and background slope are free. The tail power $n_{\rm CB}$ remains fixed.

In this example, the component lies $25$~MeV below the $\Upsilon(1S)$, within a $70$~MeV mass resolution. Fig.~\ref{fig:mass_spectrum} shows that a single-peak fit largely absorbs it, with a fitted location shift $\delta m_{\rm fit}\equiv m_{\rm fit}-m_{\Q}\simeq-1.2$~MeV. This fitted location is distinct from the calibrated mean in Eq.~\eqref{eq:moment_yield}.

\subsection{Change in nuclear observables}
At fixed collision energy and rapidity acceptance, we follow the same $r_{pp}(p_T)$ into $pA$ and $AA$. In the example, $X$ follows binary-collision scaling and has no elliptic flow, while quarkonium is suppressed. The fixed nuclear responses in Table~\ref{tab:inputs} determine the three observables in Fig.~\ref{fig:medium_response} through Eqs.~\eqref{eq:r_mix} and \eqref{eq:v_mix}. The chosen nonzero $v_2^{\Q}$ illustrates dilution; the cited CMS $\Upsilon$ measurements are compatible with zero \cite{CMSUpsilonV2}.

At the reference momentum, the fraction of $X$ increases from five percent in $pp$ to about thirteen percent in $AA$. This same enrichment raises the nuclear modification factor of the total signal to $R_{AA}^{\mathrm{tot}}=0.3825$ and reduces its anisotropy to $v_2^{\mathrm{tot}}/v_2^{\Q}\simeq0.869$. These values follow from the specified component fraction and nuclear responses. The ratio in panel (c) assumes nonzero $v_2^{\Q}$; if both components have zero flow, this observable supplies no discrimination.

The calibrated mean also responds to this change in composition. For the mass splitting in Table~\ref{tab:inputs}, $\delta_{pp}=1.25$~MeV and $\delta_{AA}\simeq3.27$~MeV. Together with the nuclear modification factor above, they satisfy Eq.~\eqref{eq:moment_yield}. These mean shifts follow from Eq.~\eqref{eq:moment}, independently of the single-peak fit location.

\subsection{Dependence on mass separation and resolution}
Fig.~\ref{fig:mass_contours} extends the mass fit to both signs of $\Delta m$ and a range of resolutions. Positive $\Delta m$ places $X$ below the quarkonium peak. The one-sided tail and falling background make the fitted displacement asymmetric. The contours show how the absorption into a single-peak fit depends on mass separation and resolution, with the remaining inputs held fixed.

\section{Discussion}
A joint likelihood would combine production, mass, and decay-angle distributions with nuclear yields and flow under one candidate hypothesis. Mass information can constrain the mixture, while nuclear observables test its specified response. The likelihood must include uncertainties in quarkonium production and nuclear effects \cite{Andronic2016,AndronicArnaldi2025}, correlated detector and normalization uncertainties, and independent constraints. Signal-extraction biases must be calibrated under the candidate hypothesis, and uncertainty in an energy-interpolated $pp$ reference propagated. The search tests whether a candidate is required beyond the allowed QCD and detector variations; residual tensions also require testing unmodelled Standard Model sources.

Proton--nucleus data help constrain modifications of the incoming parton distributions and other small-system effects. Decay polarization supplies further information through the muon decay-angle distribution folded with acceptance and efficiency; its fitted coefficients are not simple yield-weighted averages. Dedicated resonance searches and other decay channels constrain the same candidate hypothesis.

Equal component modification factors in a selected bin leave the composition unchanged; exact mass degeneracy removes the first-moment test. Enrichment alone does not demonstrate a gain over $pp$ searches. Such a gain depends on event counts, backgrounds, mass resolution, and correlated uncertainties. The examples quantify changes at fixed inputs, not $pp$-only versus joint search sensitivities.

\balance
\section{Conclusions}
We have proposed testing new contributions hidden within established signals by linking particle-physics searches with heavy-ion measurements of the same final state. A common candidate hypothesis specifies how the component contributes to each system. The $\Upsilon(1S)$ illustration shows that a contribution absorbed in $pp$ fits can produce correlated changes in nuclear yields, mass moments, and elliptic flow when its nuclear response differs from that of quarkonium.

Under the stated mass-response conditions, $R_s^{\mathrm{tot}}\delta_s=R_s^X\delta_{pp}$ eliminates the unknown quarkonium yield suppression. For independently constrained $R_s^X$, it tests a candidate against calibrated mass moments and total yields. Its application requires control of the quarkonium mass profile and detector uncertainties; its statistical value depends on the data samples. Joint $pp$, $pA$, and $AA$ measurements thus confront one candidate with distinct nuclear environments and can expose contributions absorbed in separate interpretations.
\label{sec:conclusions}
\FloatBarrier

\section*{Statements and declarations}
\noindent\textbf{Funding.} This work was supported by Academia Sinica, National Cheng Kung University, and the National Science and Technology Council of Taiwan.

\noindent\textbf{Competing interests.} The author declares no competing interests.

\noindent\textbf{Data availability.} The numerical examples are generated from the response models specified in the text. The accompanying Python code reproduces the figures.

\noindent\textbf{Use of AI-assisted tools.} The author used ChatGPT (OpenAI) for language and organizational editing, literature searches, code preparation, and assistance with algebraic derivations and checks. The author reviewed and edited the content and takes responsibility for the scientific ideas, calculations, interpretation, references, and final manuscript.

\end{document}